# Compact localized boundary states in a quasi-1D electronic diamond-necklace chain

S. N. Kempkes,[1, *] P. Capiod,[2, 3, *] S. Ismaili,[1] J. Mulkens,[2] L. Eek[1], I. Swart,[2, †] and C. Morais Smith[1, ‡]

[1]*Institute for Theoretical Physics, Utrecht University, Utrecht, Netherlands*
[2]*Debye Institute for Nanomaterials Science, Utrecht University, Utrecht, Netherlands*
[3]*Univ. Lille, CNRS, Centrale Lille, Junia, Univ. Polytechnique Hauts-de-France, UMR 8520 - IEMN – Institut d'Electronique de Microélectronique et de Nanotechnologie, F-59000 Lille, France*
(Dated: November 18, 2022)

## Abstract

**Zero-energy modes localized at the ends of one-dimensional (1D) wires hold great potential as qubits for fault-tolerant quantum computing. However, all the candidates known to date exhibit a wave function that decays exponentially into the bulk and hybridizes with other nearby zero-modes, thus hampering their use for braiding operations. Here, we show that a quasi-1D diamond-necklace chain exhibits an unforeseen type of robust boundary state, namely compact localized zero- energy modes that do not decay into the bulk. We find that this state emerges due to the presence of a latent symmetry in the system. We experimentally realize the diamond-necklace chain in an electronic quantum simulator setup.**

---

* Both authors contributed equally.
† Correspondence to: i.swart@uu.nl
‡ Correspondence to: C.deMoraisSmith@uu.nl

Topological states of matter attracted a lot of interest in previous years because of their potential use as qubits in a quantum computer[1–4]. One of the difficulties concerning quantum computing with topological states such as the non-Abelian Majorana bound states in a Kitaev chain[1,4,5] is their exponential decay into the bulk. When a Kitaev chain is too short, the quasiparticle Majorana-bound states at both edges hybridize and move away from zero energy. Therefore, in order to have a proper quantum computation, the length of the chain $L$ should be long in comparison with the characteristic coherence length $\xi$, such that the amplitude of the exponentially decaying wave function ($\propto \exp(-L/\xi)$) is small at the other side of the chain. The exponential decay of electronic states is also a problem for other systems, e.g. hybridization was experimentally shown to be an important factor in a quantum gate device[5] and for the poor man's quantum gate based on 0D boundary modes in the SSH model.[6] It would therefore be beneficial to have robust states that are fully localized on the ends of a 1D wire.

Fully localized states are known to exist in the *bulk* of certain lattices[7–16]. These states correspond to eigenstates of the Hamiltonian that are completely localized in a certain sub-region of the lattice and have strictly zero amplitude otherwise. Due to their local character, these compact localized states are protected against perturbations outside the sites where they are located[13]. They occur in crystalline flat-band systems, where often frustration is causing the modes to be completely localized. As a result, these compact localized states do not mix with other bulk states and can be excited in a relatively easy manner, as shown experimentally in Refs.[11,12,17]. Furthermore, it has been recently proposed that these states could be used in a quantum network to transfer information in a proper and experimentally feasible manner[15]. Examples of fully localized states can be found in (quasi)-1D lattices such as diamond, stub or cross chains, and in 2D, such as the Lieb lattice [8,9,11,12,16,18–21].

Here, we propose a model, namely non-interacting electrons in a *quasi-1D diamond-necklace chain*, for which robust compact localized states occur at the *edges*. This chain bears some resemblance with the diamond chain, although in the latter the compact localized states are bulk modes[10,13,22–24]. The diamond-necklace chain has been studied in the context of spin chains[25–27], where it is known as the dimer-plaquette chain, and recently in the context of flat bands in a non-interacting lattice[28]. The end modes that we find are doubly degenerate, have an energy in the insulating bulk gap, are compactly localized at the extremities of the lattice (no bulk decay) and are robust against a large number of perturbations. Furthermore, we show that the amplitude of the wave function of the zero mode can be fully controlled via either introducing anisotropies in the hopping amplitude or a flux in the plaquettes. We examine these compact states in an experimental setup and verify the theoretical proposal of compact localized end states in the diamond-necklace chain. These states open the path to the manipulation of boundary zero modes without the problem of hybridization of the end modes.

The experimental setup that we use to verify the theoretical calculations is based on the electronic quantum simulator using CO adsorbed on a Cu(111) sample[29–32]. A Cu(111) substrate exhibits a 2D electron gas at its surface. The CO molecules act as repulsive scatterers for the surface electrons of the Cu(111) substrate, confining them to the area between the CO molecules[30,31,33,34]. This method and similar ones have been successfully used to fabricate flat-band models such as the Lieb lattice [31,35] and stub, diamond and cross lattices[16]. Further, these setups have been used to show robust zero modes in an SSH model[35], in a 2D kagome[34] and in a kekulé lattice[33]. However, in all these previous examples the corner modes decay exponentially into the bulk, contrarily to the modes identified here. The experimental results are compared with tight-binding and muffin-tin calculations, see methods for further details. In the remainder, we first discuss the diamond-necklace chain in more detail and then describe the experiment.

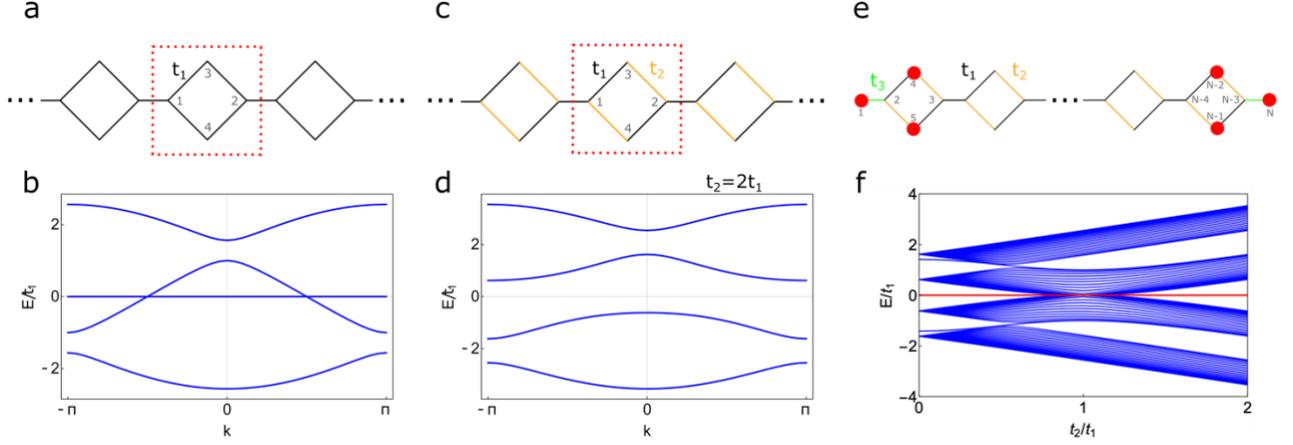

Figure 1. The quasi-1D diamond-necklace chain. **(a)** Periodic diamond-necklace chain with hopping $t_1$ connecting the four lattice sites (numbering indicated in grey). **(b)** Band structure for the lattice shown in (a). The band structure shows a flat band at $E = 0$. **(c)** The diamond-necklace lattice with the hopping $t_2$ between sites 1-4 and 2-3, and $t_1$ otherwise. **(d)** Band structure for the lattice show in (c) with $t_2 = 2t_1$. When $t_2 \neq t_1$, a band gap opens up at $E = 0$. **(e)** Finite- size lattice ending on both sides with a bond $t_3$. In this case, one can always find a degenerate state with $E = 0$ that is compactly localized at the boundaries of the chain when $t_2 \neq t_1$. The amplitudes of these wave-functions are schematically shown in red. They depend on the hopping parameters $t_1$, $t_2$ and $t_3$. **(f)** Spectrum of the finite-size lattice shown in (e), consisting of $N = 82$ sites and hopping parameters $t_3 = t_1$. The zero-mode localized at the end of the chain is shown in red and is always compactly localized (no bulk decay) in the three lattice sites at the boundaries when $t_2 \neq t_1$.

The quasi-1D diamond-necklace lattice is shown in Fig. 1a. The lattice consists of 4 sites in a unit cell, connected with a hopping $t_1$. The Bloch Hamiltonian is given by

$$H(k) = \begin{pmatrix} \epsilon & -t_1 e^{-ik} & -t_1 & -t_1 \\ -t_1 e^{-ik} & \epsilon & -t_1 & -t_1 \\ -t_1 & -t_1 & \epsilon & 0 \\ -t_1 & -t_1 & 0 & \epsilon \end{pmatrix} \qquad (1)$$

where $k$ is the wave number and $\epsilon$ the onsite energy. Apart from three dispersive bands, the spectrum shows a flat band at energy $E = \epsilon$ corresponding to a wave function $|\psi\rangle = (0,0,1,-1)^T$, which is completely localized on the sites 3 and 4. The spectrum with $\epsilon = 0$ is shown in Fig. 1b. Now, we can open a gap in the spectrum by introducing a hopping $t_2 \neq t_1$ between sites 1-4 and 2-3, as shown in Figs. 1c-d. The localized state is no longer a solution to the Schrödinger equation and there is a gap at $E = \epsilon = 0$.

In a finite chain consisting of $N$ sites, this bandgap opening gives rise to compact localized boundary states. The finite chain is shown in Fig. 1e; the chain starts and ends with a hopping $t_3$, which allows us to tune the amplitude of the localized wave functions. The spectrum of the finite chain as a function of $t_2/t_1$ is shown in Fig. 1f. In this finite chain, there is a zero-energy end mode, indicated in red in the spectrum. These states are compactly localized on sites 1, 4 and 5 on the left side of the chain and on sites $N-2, N-1$ and $N$ on the right side of the chain when $t_2 \neq t_1$, as schematically shown with red disks in Fig. 1e. The modes can be understood as a hybrid of the compact localized bulk states in a diamond chain[10,13,36] and a boundary mode in the SSH model[37]. When considering the limit $t_3 = 0$, there are two isolated sites on either side of the chain, with a localized wave function at energy $E = \epsilon$. If $t_1 = t_2$, there are compact states in the bulk and the end mode can hybridize with the compact states near the boundary. When $t_1 \neq t_2$, there is a gap in the spectrum and therefore no state with the same energy in the bulk to hybridize with the edge mode. In this sense, one could expect an exponential decay from the end-localized states into the bulk if the hopping $t_3 \neq 0$, in a similar way as it occurs in the SSH model[37]. However, due to destructive interference, the zero mode does not decay exponentially into the bulk but remains compactly localized at the edges.

We can write down an exact form of the wave function by making use of destructive interference [38–40]. We are looking for a (not-normalized) wave function of the form $|\psi\rangle = (1,0,0,r_1,r_2,0,0,\ldots)^T$ that has only an amplitude on the sites 1, 4 and 5 and energy $E = \epsilon$. When acting on our trial wave function with the Hamiltonian corresponding to the finite chain, we find:

$$H|\psi\rangle = \begin{pmatrix} \epsilon & -t_3 & 0 & 0 & 0 & 0 & \cdots \\ -t_3 & \epsilon & 0 & -t_1 & -t_2 & 0 & \cdots \\ 0 & 0 & \epsilon & -t_2 & -t_1 & -t_1 & \cdots \\ 0 & -t_1 & -t_2 & \epsilon & 0 & 0 & \cdots \\ 0 & -t_2 & -t_1 & 0 & \epsilon & 0 & \cdots \\ 0 & 0 & -t_1 & 0 & 0 & \epsilon & \cdots \\ \vdots & \vdots & \vdots & \vdots & \vdots & \vdots & \ddots \end{pmatrix} \begin{pmatrix} 1 \\ 0 \\ 0 \\ r_1 \\ r_2 \\ 0 \\ \vdots \end{pmatrix}$$

$$= \epsilon \begin{pmatrix} 1 \\ -(t_3 + t_1 r_1 + t_2 r_2)/\epsilon \\ -(t_2 r_1 + t_1 r_2)/\epsilon \\ r_1 \\ r_2 \\ 0 \\ \vdots \end{pmatrix}. \tag{2}$$

The wave function is a solution to the Schrödinger equation when $t_3 + t_1 r_1 + t_2 r_2 = 0$ and $t_2 r_1 + t_1 r_2 = 0$, which gives $r_1 = t_3 t_1/t$ and $r_2 = t_3 t_2/t$, where $t = t_2^2 - t_1^2$. The eigenfunction with energy $E$ is then given by $|\psi\rangle = (1,0,0,t_3 t_1/t, t_3 t_2/t, 0,0,\ldots)^T$ and is completely localized. A similar calculation holds for the wave function localized on the right side of the chain. When $t_2 = t_1$, the amplitude on site 1 becomes zero and we obtain the compact localized state for the sites 4 and 5 corresponding to the states in the bulk flat band shown in Fig. 1b.

Another way to open the bulk gap in the spectrum is to introduce a flux in the diamond part of the necklace chain, as it was experimentally realized for a diamond chain[13]. The Bloch Hamiltonian for the quasi-1D diamond necklace chain is given by

$$H(k) = \begin{pmatrix} \epsilon & t_1 e^{-ik} & -t_1 e^{-i\varphi} & -t_1 \\ -t_1 e^{-ik} & \epsilon & -t_1 & -t_1 \\ -t_1 e^{-i\varphi} & -t_1 & \epsilon & 0 \\ -t_1 & t_1 & 0 & \epsilon \end{pmatrix}$$

where $t$ is the hopping amplitude, $k$ the wave number, $\epsilon$ the onsite energy, and $\varphi$ the flux per diamond. Introducing a nonzero flux in the diamond-necklace chain opens a band gap at $E = \epsilon = 0$, see Figures 2a-d, similar to the anisotropic hopping described above. A π-flux gives rise to flat bands, in the same way as the Aharonov-Bohm cages do in the diamond chain[13], see Fig. 2d. In a finite-size lattice, a non-zero flux immediately gives rise to compact states as well. Using the same wave function as above, $|\psi\rangle = (1,0,0,r_1,r_2,0,0,\ldots)^T$, we find $r_1 = 1/[1 - \exp(i\varphi)]$ and $r_2 = -r_1$ for a compactly localized state, see Figs. 2e-f.

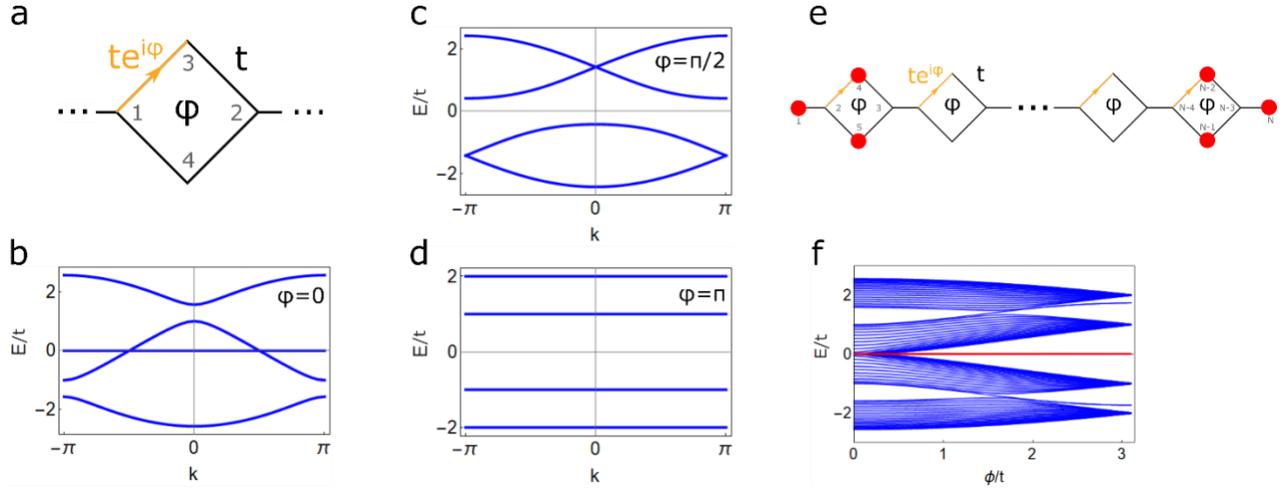

Figure 2. Flux in the diamond-necklace chain. **(a)** Unit cell of the diamond-necklace chain with hopping $t$ and flux $\varphi$. **(b)** Band structure with $\varphi = 0$ and $\epsilon = 0$. **(c)** Band structure with $\varphi = \pi/2$. A gap opens up at $E = 0$. **(d)** Band structure with a $\pi$-flux. All the bands are completely flat. **(e)** Schematic of a finite-size lattice with a flux. The compact localized states are indicated by the red circles. **(f)** Band structure of a finite chain consisting of $N = 82$ sites. A non-zero flux opens up a gap at zero energy and gives rise to states compactly localized on sites 1, 4 and 5, and N-2, N-1 and N.

Now, we examine some particular properties of these end modes. Since these modes are compactly localized, any perturbation outside of the boundary region will not disturb them. More generally, these modes are protected against any perturbation that does not couple to the sites 1, 4 and 5, and perturbations that preserve the destructive interference when connecting to sites 1, 4 and 5.

We consider two different types of perturbations, on-site and higher-order hopping. Since the compact localized state resides on 3 sites, it is unaffected by on-site disorder at different sites. This may be seen in Fig. 3a, which shows the spectrum for on-site disorder at site 2: the zero-energy mode remains intact. On the other hand, Fig. 3b depicts the spectrum for on-site disorder at site 4, which breaks the compact localized state. Finally, Fig. 3c shows the spectrum for on-site disorder sitting in the bulk cells of the diamond-necklace chain. This type of disorder does not influence the compact localized state. Next, we analyze which perturbations are allowed to keep the boundary mode localized and pinned to zero energy. Therefore, we observe what happens with the wave function when applying the general perturbations $a, b, c, d, \ldots, o$ (other perturbations are zero) in combination with different hopping parameters ($t_1$ to $t_6$, see Fig. 3f). We find

$$H|\psi\rangle = \begin{pmatrix} \epsilon & -t_1 & a & b & c & d & e & \cdots \\ -t_1 & \epsilon & f & -t_2 & -t_3 & g & h & \cdots \\ a & f & \epsilon & -t_4 & -t_5 & -t_6 & i & \cdots \\ b & -t_2 & -t_4 & \epsilon & j & k & l & \cdots \\ c & -t_3 & -t_5 & j & \epsilon & m & n & \cdots \\ d & g & -t_6 & k & m & \epsilon & o & \cdots \\ e & h & i & l & n & o & \epsilon & \cdots \\ \vdots & \vdots & \vdots & \vdots & \vdots & \vdots & \vdots & \ddots \end{pmatrix} \begin{pmatrix} 1 \\ 0 \\ 0 \\ r_1 \\ r_2 \\ 0 \\ 0 \\ \vdots \end{pmatrix} = \begin{pmatrix} \epsilon + br_1 + cr_2 \\ -t_1 - t_2 r_1 - t_3 r_2 \\ a - t_4 r_1 - t_5 r_2 \\ b + \epsilon r_1 + jr_2 \\ c + jr_1 + \epsilon r_2 \\ d + kr_1 + mr_2 \\ e + lr_1 + nr_2 \\ \vdots \end{pmatrix}. \quad (3)$$

From the latter expression, we observe that the wave-function amplitude on sites 1, 4 and 5 depends on the perturbing constants b, c, and j. To find the solution that obeys the Schrödinger equation with this eigenstate and energy $\epsilon$, we need to solve 7 equations simultaneously (one for each line). There is no general solution for these equations. To simplify the problem, we set the constants that perturb the sites 1,4 and 5 to zero, i.e. $b = c = j = 0$. Further, there is no general solution when the equations in the last two lines are present in the general form $d + kr_1 + mr_2$ and $e + lr_1 + nr_2$. We therefore set those parameters $d, k, m, e, l$ and $n$ to zero as well, such that we have

$$H|\psi\rangle = \begin{pmatrix} \epsilon & -t_1 & a & 0 & 0 & 0 & 0 & \cdots \\ -t_1 & \epsilon & f & -t_2 & -t_3 & g & h & \cdots \\ a & f & \epsilon & -t_4 & -t_5 & -t_6 & i & \cdots \\ 0 & -t_2 & -t_4 & \epsilon & 0 & 0 & 0 & \cdots \\ 0 & -t_3 & -t_5 & 0 & \epsilon & 0 & 0 & \cdots \\ 0 & g & -t_6 & 0 & 0 & \epsilon & o & \cdots \\ 0 & h & i & 0 & 0 & o & \epsilon & \cdots \\ \vdots & \vdots & \vdots & \vdots & \vdots & \vdots & \vdots & \ddots \end{pmatrix} \begin{pmatrix} 1 \\ 0 \\ 0 \\ r_1 \\ r_2 \\ 0 \\ 0 \\ \vdots \end{pmatrix}$$

$$= \begin{pmatrix} \epsilon \\ -t_1 - t_2 r_1 - t_3 r_2 \\ a - t_4 r_1 - t_5 r_2 \\ \epsilon r_1 \\ \epsilon r_2 \\ 0 \\ 0 \\ \vdots \end{pmatrix} = \epsilon \begin{pmatrix} 1 \\ 0 \\ 0 \\ \frac{at_3 - t_1 t_5}{t_2 t_5 - t_3 t_4} \\ \frac{at_2 - t_1 t_4}{t_3 t_4 - t_2 t_5} \\ 0 \\ 0 \\ \vdots \end{pmatrix}, \qquad (4)$$

where the values for $r_1 = (at_3 - t_1 t_5)/(t_2 t_5 - t_3 t_4)$ and $r_2 = (at_2 - t_1 t_4)/(t_3 t_4 - t_2 t_5)$ were substituted into the last equality. In this way, we find an analytic expression for the compact boundary states. These couplings and allowed perturbations $a, f$ and $g$ are schematically shown in Fig. 3f. We further note that other perturbations are allowed if the destructive interference is preserved. For example, take line 6 in Eq. (3): $d + kr_1 + mr_2 = 0$. This line corresponds to connecting sites 1, 4 and 5 to site 6. From the analysis of the perturbation, we know that $r_2/r_1 = -(at_2 - t_1 t_4)/(at_3 - t_1 t_5) = A$. When the constants are chosen such that $d = -(k + mA)r_1$, these perturbations will not affect the compact localized state. A similar analysis leads to $e = -(l + nA)r_1$ for the perturbation in line 7 of Eq. (3), where sites 1, 4 and 5 are connected to site 7, and similar expressions follow in general for all sites connecting to the sites 1, 4, and 5. Furthermore, all perturbations that do not couple to the sites 1, 4 and 5 are allowed trivially. The same analysis can be done for the compact state localized on the right side of the chain. Therefore, we conclude that these modes are robust against many perturbations. We do however note that because of the perturbation, the bulk spectrum may be deformed, such that the zero modes are no longer gapped out.

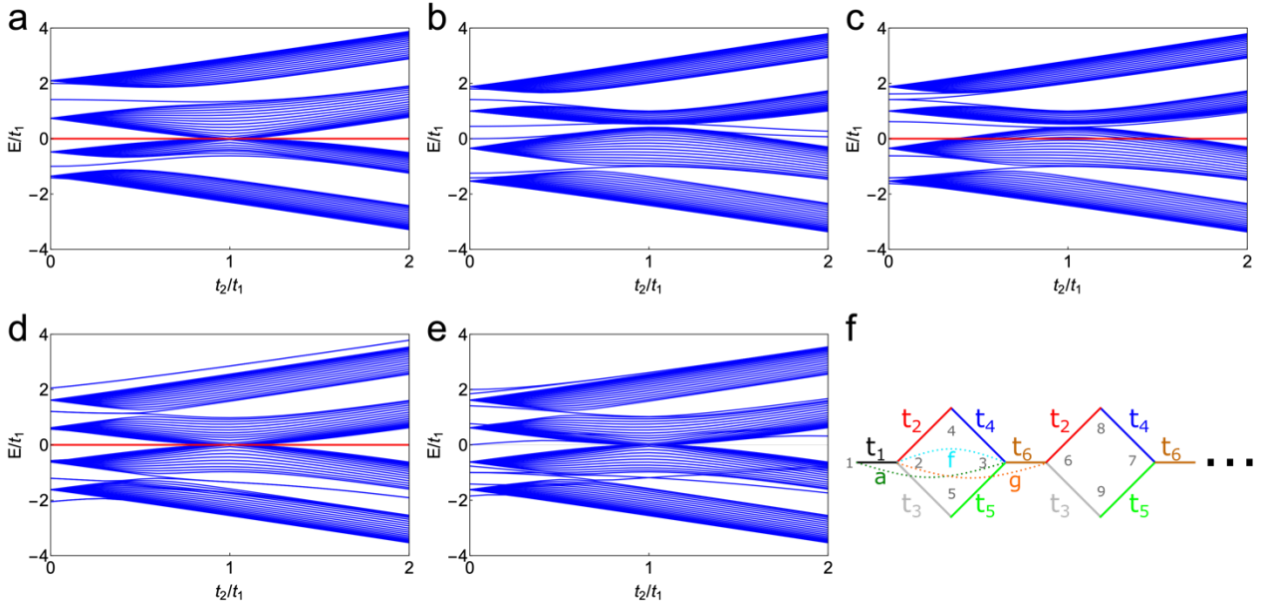

Fig. 3. Spectrum of the diamond-necklace chain shown in Fig. 1f upon inclusion of perturbations and for $t_3 = 1$. **(a)** Spectrum for an on-site perturbation with strength $v = 1$ at site 2 (and similarly at the other end of the chain). The compact localized state is unaffected. **(b)** On-site perturbation with strength $v = 1$ at site 4 (and similarly at the other end of the chain), the compact localized state vanishes. **(c)** On-site bulk perturbation with strength $v = 1$ at sites 8,12,16…, which leave the compact localized state unaffected. **(d)** General hopping perturbation $f = 1$, which preserves the compact localized state. **(e)** General hopping perturbation $b = 1$, which destroys the compact localized state. **(f)** Schematic of the allowed perturbations $a$, $f$ and $g$ with general hopping parameters in the diamond-necklace chain (perturbations $h$ and $i$ are not included in this image to prevent clumpering).

Their compact nature is what makes these end modes robust against any kind of disorder in the bulk of the crystalline lattice. Since the end-localized zero mode has no exponential decay into the bulk, these states cannot hybridize with each other and gap out. Therefore, these states are not affected by finite-size effects, and the states remain strictly at zero energy for all chain lengths. Finally, we note that the wave function amplitude at sites 1, 4 and 5 can be tuned at will. The amplitudes on sites 4 and 5 only depend on the strength of the hopping parameters $t_1$, $t_2$, and $t_3$ at the boundaries, and are not influenced by variations of the parameters in the remainder of the chain.

We verified that the robust nature of the compact localized boundary state is not a result of a symmetry protected topological phase. Rather, it is a consequence of latent symmetry. Latent symmetry is intimately connected to a symmetry between possible paths on a graph (walks of a particle along the different sites in a lattice). Any free Hamiltonian may be partitioned in $S$ and its complement $\bar{S}$

$$H = \begin{pmatrix} H_{SS} & H_{S\bar{S}} \\ H_{\bar{S}S} & H_{\bar{S}\bar{S}} \end{pmatrix}.$$

If a system has a latent symmetry, there exist a matrix $T$ such that

$$[R_S(H,E),T] = [(H^k)_{SS},T] = 0 \quad \forall k \in \{1,\ldots,N\},$$

where we introduced the isospectral reduction of $H$:[41]

$$R_S(H,E) = H_{SS} + H_{S\bar{S}}(E - H_{\bar{S}\bar{S}})^{-1}H_{\bar{S}S},$$

which in the context of condensed-matter physics is better known as an effective Hamiltonian for the $S$ degrees of freedom. In the case of the diamond-necklace chain, the choice of this partition is $S = \{1,4,5\}$, *i.e.* the sites where the compact localized state has a non-zero amplitude. Choosing the sites at which the second compact mode is localized yields a similar result. In the case of general hopping parameters $t_1 - t_6$, such as the Hamiltonian given in Eq. 4 (but without perturbations), the symmetry $T$ is given by

$$T = (|\Psi_{CLS}\rangle\langle\Psi_{CLS}|)_{SS} = \begin{pmatrix} 1 & \frac{-t_1 t_5}{t_2 t_5 - t_3 t_4} & \frac{t_1 t_4}{t_2 t_5 - t_3 t_4} \\ \frac{-t_1 t_5}{t_2 t_5 - t_3 t_4} & \frac{t_2^2 t_5^2}{(t_2 t_5 - t_3 t_4)^2} & \frac{-t_1^2 t_4 t_5}{(t_2 t_5 - t_3 t_4)^2} \\ \frac{t_1 t_4}{t_2 t_5 - t_3 t_4} & \frac{-t_1^2 t_4 t_5}{(t_2 t_5 - t_3 t_4)^2} & \frac{t_1^2 t_4^2}{(t_2 t_5 - t_3 t_4)^2} \end{pmatrix},$$

which does indeed commute with $R_S(H, E)$. Moreover, we may introduce a matrix $Q = T \oplus I_{\bar{SS}}$, with $I_{\bar{SS}}$ the identity operator on the $\bar{S}$ degrees of freedom, that $[H, Q] = 0$. The presence of a latent symmetry is in one-to-one correspondence with the existence of compact localized states. Consequently, any perturbation conserving latent symmetries preserves the compact localized boundary state.[14,15,42] Much like topologically protected phases, the edge states described here remain for a wide range of parameter choices and are robust against any type of disorder respecting the symmetry at hand.

The theory presented is now confronted with experiment. Figure 4a shows a constant-current image of a diamond-necklace chain realized by positioning CO molecules on a Cu(111) surface using the tip of an STM (ScientaOmicron LT-STM) operating at $T$ = 4K. Each CO molecule (black contrast) has been moved individually and is positioned using a procedure described in the literature[43–45]. The confined regions define atomic sites, see Fig. 4a. The boundary hopping $t_3$ is controlled by positioning the highlighted COs (black dots surrounded by white circles) as shown in Figs. 4b-d. In Fig. 4b, the highlighted CO molecules are far away from each other (1.28 nm), leading to a strong coupling between the neighboring sites (i.e. large $t_3$). In contrast, the highlighted COs are closer to each other (1.024 nm) in Fig. 4d, which decreases the boundary hopping amplitude $t_3$. The experimental spectra corresponding to the LDOS for the sites indicated in Figs. 4b-d are given in Figs. 4e-g (solid lines). Note that we only show the spectra on the left side of the chain, since the spectra on the right are similar by rotational symmetry.

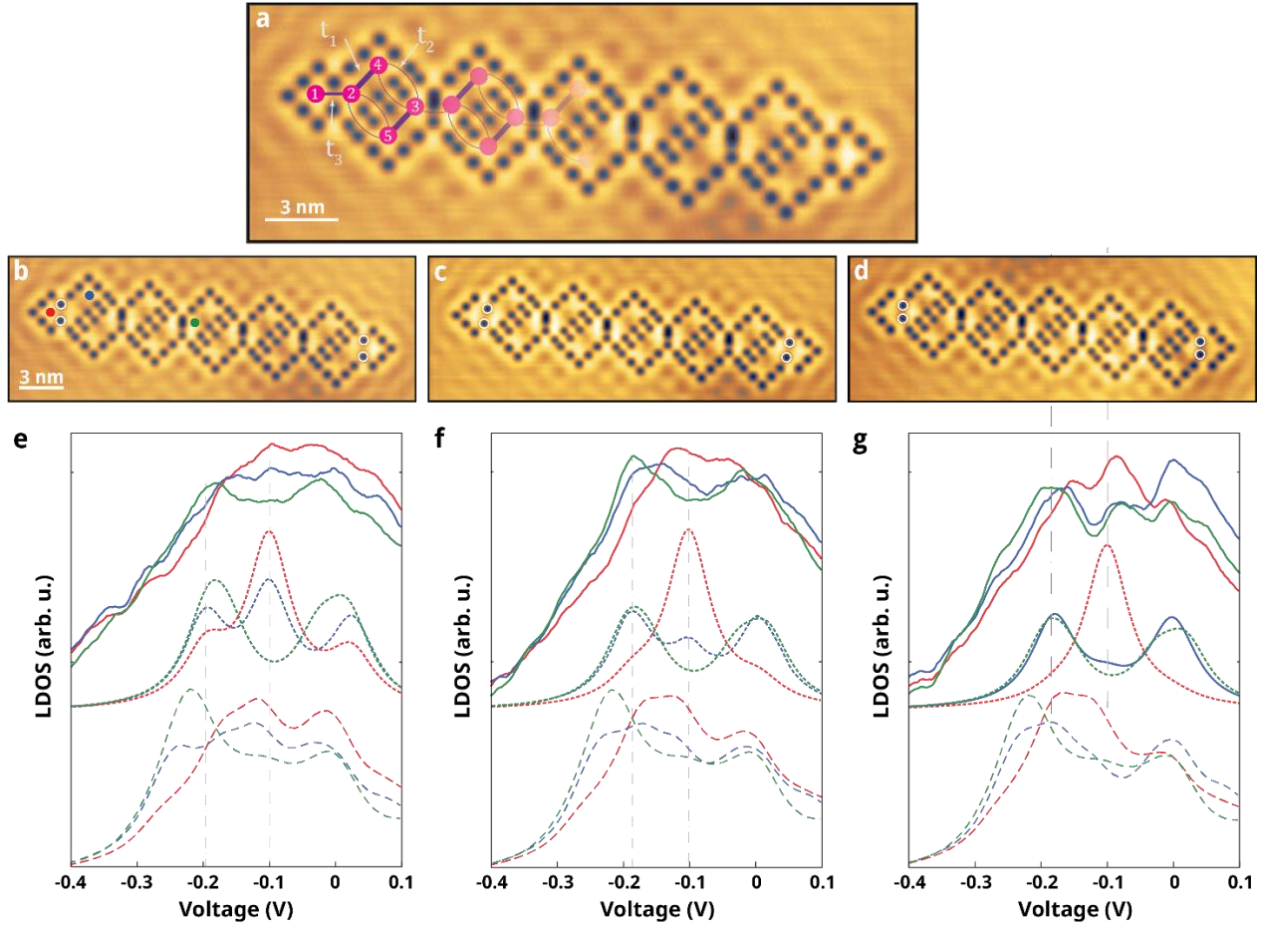

Figure 4. Experimental configuration and LDOS of the quasi-1D diamond-necklace chain. **(a)** Constant-current STM image of the diamond-necklace chain realized with CO molecules on a Cu(111) surface. Artificial atom sites and hopping terms are indicated by pink and purple circles and lines respectively. Thicker (thinner) lines represent stronger (weaker) hopping. **(b-d)** STM images of the diamond-necklace chain with a strong (a), intermediate (b) and weak (c) hopping $t_3$. The CO adsorbates are shown in black, of which four COs are highlighted. The highlighted COs determine the strength of the boundary hopping $t_3$. **(e-g)** Experimental spectra (solid lines) compared to the tight-binding (dotted lines) and muffin-tin (dashed lines) spectra of the sites indicated in (b-d), for the strong (e), intermediate (f), and weak (g) hopping parameters, respectively. Here, $t_1$ = 0.095 eV, $t_2$ = 0.1$t_1$, and $t_3$ goes from 0.8$t_1$ (e), to 0.5$t_1$ (f) and 0.3$t_1$ (g).

Upon inspection, we observe that the red spectrum (sites 1 and $N$, respectively) always has a peak-like structure around the onsite energy $V = -0.1$ V, whereas the bulk sites exhibit a gap-like structure around that energy (green site in Fig. 4b). We note that the intensity of the compact localized state is lower in experiment and muffin-tin simulations than in the tight-binding results. We tentatively attribute this to a non-negligible next-nearest-neighbor coupling (not taken into account in the tight-binding calculations). Broadening, due to scattering of surface state electrons by the CO molecules, leads to a less well-developed gap in the experimental data. However, all qualitative features of the tight-binding model are observed in the experimental data, demonstrating the experimental realization of the diamond-necklace chain.

By positioning the highlighted CO molecules differently, one can change the LDOS of the blue site from exhibiting a peak (Fig. 4e) to having a dip (Fig. 4g) around $V = -0.1$ V. Hence, the amplitude of the wave function on that site can be modified via minor changes in the coupling strength $t_3$. The experimental observations are verified by a finite-size tight-binding and muffin-tin calculations (dotted and dashed lines in Figs. 4e-g, respectively). In addition to the strong hopping parameter $t_1$ = 0.095 eV and the weak hopping $t_2 = 0.1t_1$ presented in Fig. 1e, we introduce the hopping $t_4 = 0.4t_1$ that connects the diamonds. To make the comparison with the experimental spectra, we only change the boundary hopping parameter $t_3$ from $0.8t_1$ (e), to $0.5t_1$ (f) and $0.3t_1$ (g) (and orbital overlap in a similar way, see method section). In the tight-binding LDOS, we clearly observe a large change in the blue spectra, whereas the other spectra remain similar.

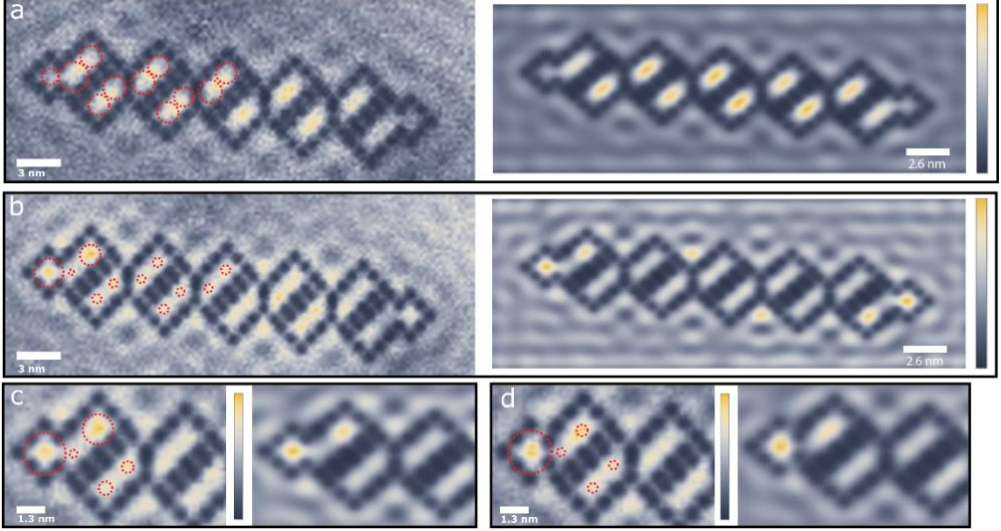

Figure 5. **(a)** Experimental (left) and muffin-tin (right) local density of states maps for the chain with strong boundary hopping at $V = -0.213$ V. The tight-binding result is indicated on top of the experimental map with red dashed circles. The radius of the circle scales linearly with $|\psi|^2$ on the indicated site. **(b)** Same as (a) but now for $V = -0.122$ V. The amplitude of the wave function is now mainly localized at the ends of the chain. We note that due to imperfections in the determination of the sample tilt, the maps show deviations between the left and right sides of the chain. **(c-d)** Zoom in on the edge of the strong (c) and the weak (d) boundary hopping chain at $V = -0.122$ V. Left and right panels correspond to experimental and muffin-tin simulated maps, respectively.

Next, we present local density of states maps of the created lattices with strong and weak coupling $t_3$ in Fig. 5. In Fig. 5a, we show the experimental (left) and muffin-tin (right) simulated maps for the strong boundary coupling chain at $V = -0.213$ V. The LDOS computed from tight binding is represented as circles on top of the experimental data, where the circle radius scales linearly with $|\psi|^2$. At this energy, the electronic LDOS is mainly localized in the bulk of the chain (bright colors), whereas it is absent in the end sites (dark colors). When increasing the voltage to $V = -0.122$ V, the LDOS becomes more pronounced at the end sites of the quasi-1D chain, especially at sites 1 and 4. Other sites, and in particular site 2, show less intensity. A closer inspection of the end modes in the strong and weak boundary-hopping chain is shown in Figs. 5c and d, respectively. The end mode is more pronounced on site 4 (top of the first diamond) in the strong boundary bonding (Fig. 5c) configuration, and less pronounced in the weak boundary bonding (Fig. 5d). Both theoretical methods predict the same trend, c.f. left (red circles) and right panels in Figs. 5c and 5d.

To conclude, we have theoretically and experimentally introduced the notion of robust compact localized boundary states. These states are present in the insulating bulk-band gap and are completely localized at the boundary of the diamond-necklace chain. We have shown how to change the wave-function amplitude of the boundary mode by controlling the boundary-hopping parameter, both in theory and in an experiment. Since these states are doubly degenerate and do not decay into the bulk, they might be the ideal candidates for quantum operations and to store and transfer information in the same way as the topological 0D modes in an SSH chain, with the difference that the chains do not need to be long in comparison with the decoherence length of the zero modes. It would be worthwhile to investigate whether compact Majorana bound states can be realized in such a quasi-1D chain with the same non-Abelian properties as the ones in the Kitaev chain, and to perform braiding operations with those compact localized edge modes.

# METHODS

### Scanning tunneling microscopy experiments

The tunneling spectra in Fig. 4 were acquired at constant height, by placing the tip above a single site. The feedback loop is disconnected and a modulated voltage is applied to the tunneling junction. The tunneling current $I$ and conductance $dI/dV$ are measured simultaneously. The differential conductance is obtained with a lock-in amplifier (rms modulation of 10 mV at 769 Hz). All spectra were averaged using at least 18 $dI/dV$ sets of reproducible curves, followed by applying a 5-point running averaging filter. Density-of-state maps have been performed by disabling the feedback loop and activating the external voltage modulation of the lock-in. The energy has been carefully chosen from the LDOS curves (see Fig. 4), and the current has been set to 1 nA by adjusting the tip-surface distance.

### Muffin-tin simulations

The experimental platform can be simulated by describing the surface state of the Cu(111) as a 2D electron gas that is patterned with circular potential barriers (CO molecules) with a height of V = 0.9 eV and a radius R = 0.3 nm[25]. We determine the energies and wave functions of this system by numerically solving the Schrödinger equation. To account for the coupling between the surface- and bulk states of copper, a Lorentzian broadening with a FWHM of 0.08 eV is applied to the theoretically computed energy levels.

### Tight-binding calculations

A Lorentzian broadening of $\Gamma$ = 80 meV is applied to the spectra to take the scattering with the bulk states into account. Further, we solve the finite-size tight-binding model with four hopping parameters $t_1 - t_4$, as mentioned in the main text. Here, $t_1$ is the strong hopping within a diamond, $t_2$ is the weak hopping within a diamond, $t_3$ is the hopping to the boundary site and $t_4$ is the hopping connecting the diamonds. The parameters used in Figs. 4e-g are (all in eV): $e_s = -0.1$, $t_1 = 0.095$, $t_2 = 0.1 t_1$, $t_4 = 0.4 t_1$, and a nearest-neighbor orbital overlap of $s_1 = 0.1$, $s_2 = 0.1 s_1$, and $s_4 = 0.4 s_1$. Further, the hopping parameters $t_3$ (overlap $s_3$) are $t_3 = 0.8 t_1$ ($s_3 = 0.8 s_1$) in Fig. 4e, $t_3 = 0.5 t_1$ ($s_3 = 0.5 s_1$) in Fig. 4f and $t_3 = 0.3 t_1$ ($s_3 = 0.3 s_1$) in Fig. 4g.

# DECLARATIONS

### Data availability

All data underlying the results are available from the authors upon reasonable request.

### Competing interests

CMS is an editorial board member for Quantum Frontiers and was not involved in the editorial review or the decision to publish this article. All authors declare that there are no competing interests.

### Funding

The research was made possible with financial support from the European Research Council (Horizon 2020 "FRACTAL", 865570) and the Dutch Research Council (grant 16PR3245). LE, CMS and IS acknowledge the research program "Materials for the Quantum Age" (QuMat) for financial support. This program (registration number 024.005.006) is part of the Gravitation program financed by the Dutch Ministry of Education, Culture and Science (OCW).

### Author contributions

The experiments were performed by PC and JM. The calculations were performed by SNK, SI and LE. All authors contributed to analyzing the data and writing the manuscript. The research was supervised by IS and CMS.

### Acknowledgements

We are grateful to M. di Liberto for discussions and D. Vanmaekelbergh for a careful reading of the manuscript. We would also like to thank A. Moustaj and M. Röntgen for the fruitful discussions about symmetries and edge states.

---